\documentclass[preprint]{revtex4}
\usepackage{amssymb}
\usepackage{graphicx}

\begin{document}

\title{An update on the double cascade scenario in two-dimensional turbulence}

\author{G. Boffetta$^{1}$ and S. Musacchio$^{2}$}
\affiliation{$^1$Dipartimento di Fisica Generale and INFN, 
Universit\`a di Torino,
via P.Giuria 1, 10125 Torino (Italy) \\
$^{(2)}$ CNRS, Lab. J.A. Dieudonn\'e UMR 6621,
Parc Valrose, 06108 Nice (France)}

\date{\today}

\begin{abstract}
Statistical features of homogeneous, isotropic, two-dimensional turbulence
is discussed on the basis of a set of direct numerical simulations up to 
the unprecedented resolution $32768^2$.
By forcing the system at intermediate scales, narrow but clear inertial
ranges develop both for the inverse and for direct cascades
where the two Kolmogorov laws for structure functions are, 
for the first time, simultaneously observed.
The inverse cascade spectrum is found to be consistent with 
Kolmogorov-Kraichnan prediction and is robust with respect the
presence of an enstrophy flux. 
The direct cascade is found to be more sensible to finite size effects:
the exponent of the spectrum has a correction with respect theoretical
prediction which vanishes by increasing the resolution.
\end{abstract}

%\pacs{PACS?}

\maketitle

\begin{quote}
``Now, listen to me. You are living on a Plane. What you style
Flatland is the vast level surface of what I may call a fluid
on, or in, the top of which you and your countrymen move
about, without rising above it or falling below it.'' \\
{\it Flatland} by E.A. Abbott
\end{quote}

The existence of two quadratic inviscid invariants is the most
distinguishing feature of Navier Stokes equations in two dimensions.
On this basis, R.H.~Kraichnan \cite{K67} predicted many years
ago the double cascade scenario: when the turbulent flow
is sustained by an external forcing acting on a typical scale $\ell_{f}$,
an inverse cascade of kinetic energy  $E=1/2 \langle v^2 \rangle$
to large scales ($\ell \gg \ell_f$) and a direct cascade of 
enstrophy $Z=1/2 \langle \omega^2 \rangle$ to small scales
($\ell \ll \ell_f$) develop. In inverse and direct ranges of
scales the theory predicts the kinetic energy spectrum 
$E(k) \simeq \varepsilon^{2/3} k^{-5/3}$ and
$E(k) \simeq \eta^{2/3} k^{-3}$ with possible logarithmic
corrections (\cite{K71}). Here $\varepsilon$ and
$\eta \simeq k_{f}^{2} \varepsilon$ are
respectively the energy and the enstrophy injection rate.

Navier-Stokes equations in two dimensions are now the prototypical model
for turbulent systems displaying a double cascade scenario. From
two-dimensional magneto-hydro-dynamics, to many geophysical model 
(such as Charney-Hasegawa-Mima), to wave turbulence models, the 
picture originally developed by Kraichnan has found many fruitful 
applications. 

Despite the expansion of the fields of applicability, it is remarkable
that the verification of Kraichnan's theory, after more than $40$ 
years from its formulation, is still partial. 
This is due to several reasons. First of all, the difficulties
to generate a laboratory flow which is truly two dimensional on
a large range of scales, limits the experimental approaches.
From a numerical point of view, the situation
in two dimensions is apparently very convenient with respect to
three dimensions. A deeper analysis shows that this is not the case,
as the simultaneous simulation of two inertial ranges requires
very large resolutions. Moreover, because time step is proportional
to grid size, the computational effort for simulating two-dimensional 
turbulence can be even larger than in the three dimensional case.

In the present paper we report numerical results on the statistics
of the two cascades of two-dimensional turbulence on 
the basis of very high resolution (up to
$32768^2$) direct numerical simulations. 
Together with previous results at lower resolutions (already reported on
\cite{boffetta_jfm07}) 
we obtain strong indications that the classical Kraichnan scenario 
is recovered in the limit of two infinitely extended inertial ranges, 
although we are unable to address the issue of possible logarithmic 
corrections in the direct cascade. 

The motion of an incompressible (${\bf \nabla} \cdot {\bf u}=0$)  fluid 
in two dimensions is governed by the Navier--Stokes equations which
are written for the scalar vorticity field 
$\omega={\bf \nabla} \times {\bf u}$ as
\begin{equation}
\partial_t \omega + {\bf u} \cdot {\bf \nabla} \omega = \nu \nabla^2 \omega
-\alpha \omega + f_{\omega} \, .
\label{eq:1}
\end{equation}
In (\ref{eq:1}) $\nu$ is the kinematic viscosity, $f_{\omega}$ is a forcing
term and the friction term $-\alpha \omega$ removes energy at large scales
in order to reach a stationary state.
Alternatively, one can consider the quasi-stationary regime with $\alpha=0$ 
in which the integral scale grows, according
to Kolmogorov scaling, as 
$L(t)=\varepsilon^{1/2} t^{3/2}$.
In this case, Galilean invariant statistics (i.e. 
velocity structure functions or
energy spectrum) is stationary at small scales $\ell < L(t)$. 
We remark that the form of the friction term in (\ref{eq:1}) physically
represents a crude approximation of the effects induced by bottom or 
air friction on a thin layer of fluid \cite{ch_amr09}.

We numerically integrate (\ref{eq:1})
by means of a standard, fully dealiased, pseudo-spectral parallel
code on a double
periodic square domain of side $L_x=L_y=2 \pi$ at spatial resolution up to 
$N=32768$. 
The forcing term $f_{\omega}$ in (\ref{eq:1}) is $\delta$-correlated in time
(in order to control energy and enstrophy input) and peaked on a
characteristic forcing scale $\ell_f$. We use either a Gaussian
forcing with correlation function 
$\langle f_{\omega}({\bf r},t) f_{\omega}({\bf 0},0) \rangle=
F \delta(t) \exp(-(r/\ell_f)^2)$ or a forcing which has support
on a narrow band of wavenumbers around $k_f=\pi/\ell_f$ in Fourier space.
In both cases this ensures that energy and enstrophy input are localized 
in Fourier space and only a limited range of scales around the forcing scale
is affected by the details of the forcing statistics. 
More complex forcing, not localized in wavenumber
space, can have a direct effect on inertial range scales 
\cite{mmm_prl07,ch_amr09}.
The forcing scale in all run is fixed at $\ell_f=L_x/100$ to allow the 
development of inertial ranges both at scales $\ell > \ell_f$ 
(inverse cascade) and $\ell < \ell_f$ (direct cascade). 
For the largest simulation run ($N=32768$) 
we study the inverse cascade in the quasistationary regime with
$\alpha=0$ and we stop the integration of (\ref{eq:1})  when 
$L(t) < L_x$ to avoid the pile-up of energy at the largest available scale.
Table \ref{table1} reports the most important parameters for the simulations.

%%%%%%%%%%%%%%%%%%%%%%%%%%%%%%%%%%%%%%%%%%%%%%
\begin{table}
\begin{tabular}{cccccc}
Label & A & B & C & D & E \\ \hline
N & $2048$ & $4096$ & $8192$ & $16384$ & $32768$ \\ 
$\nu$ & $2\times10^{-5}$ & $5\times10^{-6}$ & $2\times10^{-6}$ &
$1\times10^{-6}$ & $2.5 \times10^{-7}$ \\
$\alpha$ & $0.015$ & $0.024$ & $0.025$ & $0.03$ & $0.0$ \\
$\ell_{f}/\ell_{\nu}$ & $13$ & $26$ & $40$ & $57$ & $116$ \\
$R_{\lambda}$ & $7.9$ & $15.4$ & $21.5$ & $26.0$ & $36.0$ \\
$\varepsilon_{\alpha}/\varepsilon_{I}$ & $0.54$ & $0.83$ & $0.92$ & $0.95$ & $0.98$ \\
$\eta_{\nu}/\eta_{I}$ & $0.96$ & $0.92$ & $0.90$ & $0.88$ & $0.98$ \\
$\delta$ & $1.8$ & $1.1$ & $0.75$ & $0.50$ & $0.35$  \\ \hline
\end{tabular}
\caption{Parameters of the simulations. $N$ spatial resolution, 
$\nu$ viscosity, $\alpha$ friction, 
$\ell_{f}=\pi/k_{f}$ forcing scale, 
$R_{\lambda}=Z^{3/2}/\eta_{\nu}$ Reynolds number for the
direct cascade \cite{hokf_jfm74},
$\ell_{\nu}=\nu^{1/2}/\eta_{\nu}^{1/6}$ enstrophy dissipative scale, 
$\varepsilon_{I}$ energy injection rate, $\varepsilon_{\alpha}$
friction energy dissipation rate (large-scale energy flux for run E),
$\eta_{I}$ enstrophy injection rate, $\eta_{\nu}$
viscous enstrophy dissipation rate, 
$\delta$ correction to the Kraichnan spectral exponent in the direct cascade.
Viscosity is tuned to 
have for all runs $k_{max} \ell_{\nu} \simeq 3$.  For the run $E$, 
for which 
$\alpha=0$, $\varepsilon_{\alpha}$ is
kinetic energy growth rate.}
\label{table1}
\end{table}
%%%%%%%%%%%%%%%%%%%%%%%%%%%%%%%%%%%%%%%%%%%%%%%%%%%%%%%%%%%%

\begin{figure}[htb!]
\includegraphics[clip=true,keepaspectratio,width=12cm]{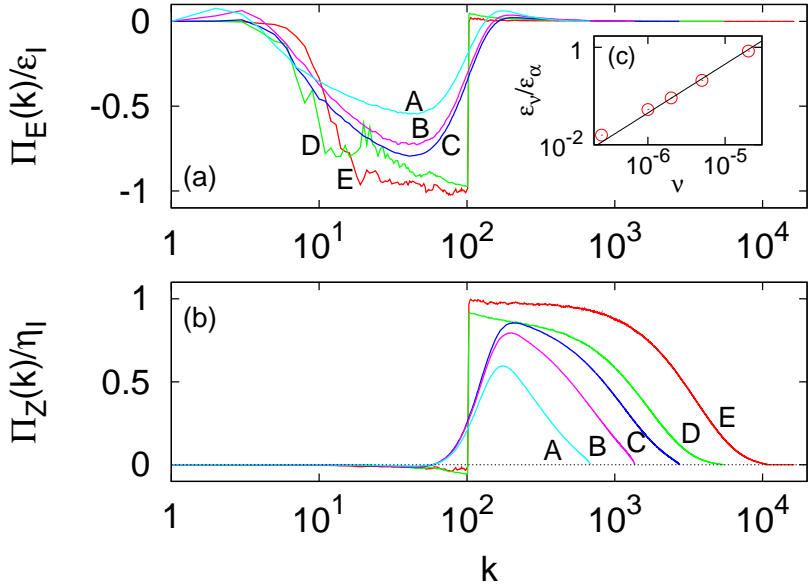}
\caption{(Color online). (a) Energy and (b) enstrophy fluxes in Fourier space 
for the runs of Table~\ref{table1}. Fluxes for runs $D$ and $E$ 
are computed from a single snapshot. Inset (c): ratio of viscous over friction
energy dissipation versus kinematic viscosity for the $5$ runs, the line
represents a linear fit.}
\label{fig1}
\end{figure}

The first information we get from the Table is related
to the direction of the energy and enstrophy fluxes.
According to the original idea of Kraichnan on the double cascade,
in the ideal case of an infinite inertial range all the energy (enstrophy) 
injected should be transferred to large (small) scales.  
This can be thought as a limit case of a realistic situation in which 
the inertial range has a finite extension because of the presence of 
large and small scale dissipation.
The characteristic viscous scale $\ell_{\nu}$ and friction scale $\ell_{\alpha}$
can be expressed in terms of the energy (enstrophy) viscous dissipation rate
$\varepsilon_{\nu}$ ($\eta_{\nu}$) and friction dissipation rate
$\varepsilon_{\alpha}$ ($\eta_{\alpha}$) by the relations 
$\ell_{\nu}^2=\varepsilon_{\nu}/\eta_{\nu}$ and 
$\ell_{\alpha}^2=\varepsilon_{\alpha}/\eta_{\alpha}$. 
Energy and enstrophy balance equations in stationary conditions give 
\cite{K67,borue93,eyink_prl95}
\begin{eqnarray}
{\varepsilon_{\nu} \over \varepsilon_{\alpha}} &=& 
\left({\ell_{\nu} \over \ell_{f}}\right)^2 
\left({\ell_{f} \over \ell_{\alpha}}\right)^2 
{(\ell_{\alpha}/\ell_{f})^2 - 1 \over 
1 - (\ell_{\nu}/\ell_{f})^2} 
\label{eq:2} \\
{\eta_{\nu} \over \eta_{\alpha}} &=&
{(\ell_{\alpha}/\ell_{f})^2 - 1 \over 
1 - (\ell_{\nu}/\ell_{f})^2}
\label{eq:3}
\end{eqnarray}
Therefore with an extended direct inertial range, 
$\ell_{\nu} \ll \ell_{f}$, one has
$\varepsilon_{\nu}/\varepsilon_{\alpha} \to 0$, i.e.
all the energy injected goes to large scales. Moreover, if 
$\ell_{\alpha} \gg \ell_{f}$ one obtains
$\eta_{\alpha}/\eta_{\nu} \to 0$ i.e. all the enstrophy
goes to small scales to generate the direct cascade.
Indeed, from Table~\ref{table1} we see that increasing the resolution,
i.e. $\ell_{f}/\ell_{\nu}$, the fraction of energy which flows to 
large scales increases. Because 
in our runs $\ell_{\alpha}/\ell_{f}$ is constant with resolution
and because $\ell_{\nu} \propto \nu^{1/2}$ we expect that,
according to (\ref{eq:2}), 
$\varepsilon_{\nu}/\varepsilon_{\alpha} \propto \nu$ \cite{K67} as indeed
is shown in the inset of Fig.~\ref{fig1}.

Most of the enstrophy (around $90 \%$) is dissipated by small 
scale viscosity. We observe a moderate increase of the large-scale
contribution to enstrophy dissipation $\eta_{\alpha}$ by going from
run $A$ to $D$. This is a finite size effect because we have to
increase the friction coefficient $\alpha$ with the resolution $N$ 
in order to keep the friction scale 
$\ell_{\alpha} \simeq \alpha^{-3/2} \varepsilon_{\alpha}^{1/2}$ constant
when $ \varepsilon_{\alpha}$ grows. Indeed, for the run $E$ 
without large-scale friction, the enstrophy flux to small scales 
almost balances the input. 

%------------------------------------------------------------------------
%------------------------------------------------------------------------

Figure~\ref{fig1} shows the energy and enstrophy fluxes in Fourier space
defined as $\Pi_{E}(k) \equiv -\partial_t \int_{0}^{k} E(k') dk'$
and $\Pi_{Z}(k) \equiv -\partial_t \int_{0}^{k} k'^2 E(k') dk'$
(where $E(k)$ is the energy spectrum and the time derivative keeps the 
contribution from nonlinear terms in (\ref{eq:1}) only \cite{frisch_95}).
We observe that, because resolution is changed by keeping 
$\ell_{\alpha} \gg \ell_{f}$ constant, the only effect of increasing
resolution on the inverse cascade is the growth of 
$\varepsilon_{\alpha}/\varepsilon_{I}$ (i.e. $\Pi_{E}(k)/\varepsilon_{I}$)
as discussed above,
while the extension of the inertial range does not change. 
Despite the limited resolution of the inverse cascade inertial range
($k_f=100$), we observe an almost constant energy flux which 
develops independently on the presence of a direct cascade inertial range 
(run A). Of course, because of the presence of the two energy sinks 
(viscosity and friction) a plateau indicating a constant energy flux 
is clearly observable for the largest resolution simulation $E$ only.
On the contrary, the direct cascade does not develop for the
small resolution runs as the dissipative scale is very close
to the forcing scale (see Table~\ref{table1}). A constant 
enstrophy flux $\Pi_{Z}(k)$ which extends over about one decade is on the
other hand obtained for the most resolved run $E$.

The behavior of the fluxes around $k \simeq k_{f}$ depends on the 
details of the injection: transition from zero to negative (positive)
energy (enstrophy) flux is sharp in the case of forcing on a narrow
band of wavenumber (run D and E) while it is more smooth for the Gaussian 
forcing which is active on more scales. 

%------------------------------------------------------------------------
\begin{figure}[htb!]
\includegraphics[clip=true,keepaspectratio,width=12cm]{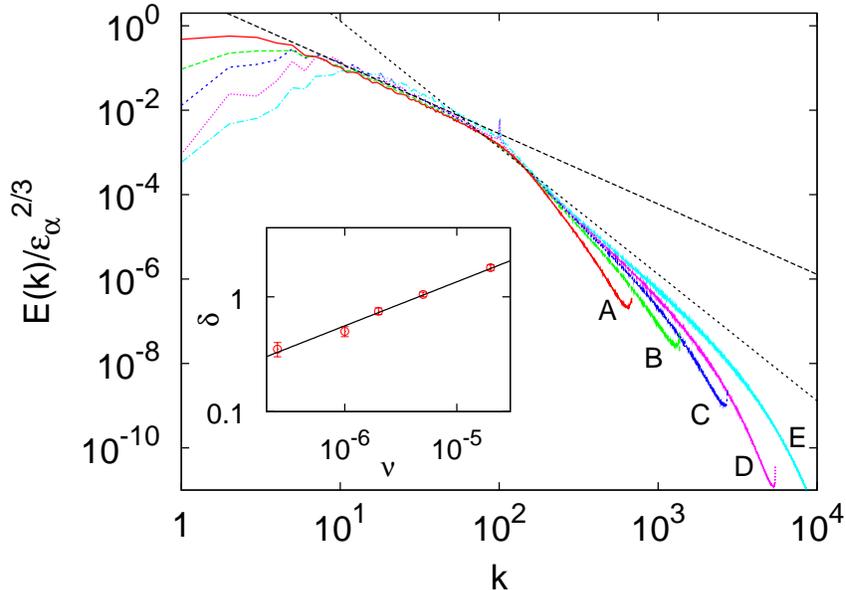}
\caption{(Color online).
Energy spectra for the simulation of Table~\ref{table1} compensated
with the inverse energy flux. Lines represent the
two Kraichnan spectra $C k^{-5/3}$ (dashed) with $C=6$ and 
$k^{-3}$ (dotted). The inset shows the correction $\delta$ to the 
Kraichnan exponent for the direct cascade $3$ obtained from the 
minimum of the local slope of the spectra in the range 
$k_f \le k \le k_{\nu}$ as a function of the viscosity. Errorbars
are obtained from the fluctuations of the local slope. The line 
has a slope $0.38$ and is a guide for the eye.}
\label{fig2}
\end{figure}
%------------------------------------------------------------------------

In Fig.~\ref{fig2} we plot the energy spectra of the different runs
compensated with the energy flux. In the inverse range 
$k<k_f$ a Kolmogorov spectrum $E(k)=C \varepsilon_{\alpha}^{2/3} k^{-5/3}$
is clearly observed for all simulations. The value of the Kolmogorov constant
$C \simeq 6$ is compatible with those obtained from more resolved
inertial range \cite{bcv_pre00} and it is found to be independent on 
the resolution.
For what concerns the direct cascade, the spectrum is steeper than 
the Kraichnan prediction $k^{-3}$. This effect is due to finite size
effects, as it reduces by increasing the resolution. In order to quantify
the recovery of the Kraichnan exponent, we computed for all
runs the local slope of the energy spectra in the range of wavenumber
$k_f \le k \le k_{\nu}$. A plateau for the slope in this range of scales
defines the scaling 
exponent $-(3+\delta)$ of the energy spectrum in the direct cascade. 
In the inset of Fig.~\ref{fig2}
we plot the measured value of the correction $\delta$ as a function of
the viscosity of the run. It is evident that, despite the fact the classical
exponent $-3$ is not observed, the indication is that it should 
eventually be recovered in the infinite
resolution limit $\nu \to 0$. It is interesting to observe that for
the most resolved run $E$, for which the enstrophy flux is almost constant
over a decade of wavenumbers (see Fig.~\ref{fig1}), the exponent of the energy
spectrum still has a significant correction $\delta \simeq 0.35$.
We remark that a clear observation of Kraichnan $k^{-3}$ spectrum
in simulations is obtained using some kind of modified viscosity
only \cite{borue93,la00,pf02}, while steeper spectra 
has also been observed in simulations of (\ref{eq:1}) with a large scale 
forcing, i.e.  resolving the direct cascade only \cite{gotoh_pre98}.
Therefore also for the direct cascade our simulations support the
picture for which the statistics of one cascade is independent
on the presence of the other cascade.

%------------------------------------------------------------------------
\begin{figure}[htb!]
\includegraphics[clip=true,keepaspectratio,width=12cm]{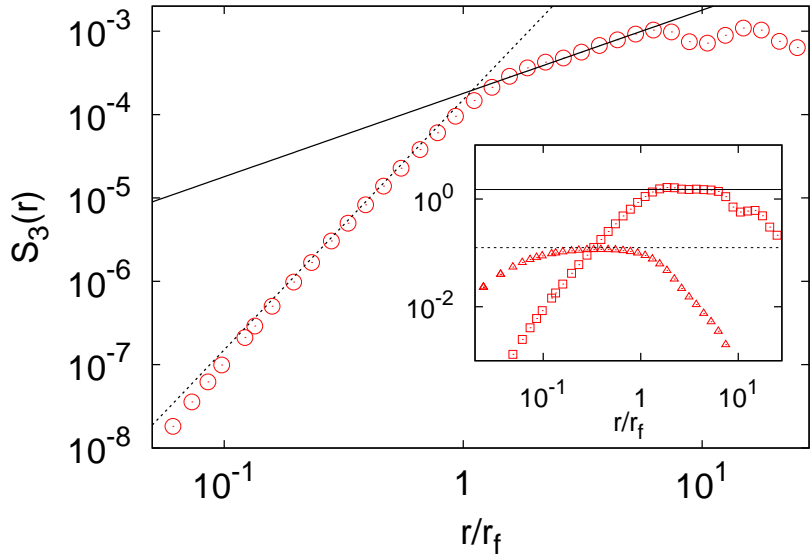}
\caption{(Color online). Third-order longitudinal velocity structure function
$S_{3}(r) \equiv \langle (\delta u_{\parallel}(r))^3 \rangle$
for run $E$ at final time. The two lines represent the Kolmogorov laws
(\ref{eq:4}) (continuous) and (\ref{eq:5}) (dotted). Inset:
compensation of $S_{3}(r)$ with $\varepsilon_{\alpha} r$ (circles)
and with $\eta_{\nu} r^3$ (triangles). Lines are the coefficient
$3/2$ (continuous) and $1/8$ (dotted).}
\label{fig3}
\end{figure}
%------------------------------------------------------------------------

We now consider small scale statistics in physical space, starting
from velocity structure function
$S_{p}(r) \equiv \langle (\delta u_{\parallel}(r))^p \rangle$
(with $(\delta u_{\parallel}(r)=({\bf u}({\bf x}+{\bf r})-{\bf u}({\bf x}))
\cdot {\bf r}/r$).
The Kolmogorov relation, a consequence of constant energy (or enstrophy)
flux in the inertial range, together with assumptions of homogeneity
and isotropy, gives an exact prediction for the third-order 
longitudinal velocity structure function $S_{3}(r)$ 
\cite{bernard_pre99,lindborg_jfm99,yakhot_pre99}. For 
the inverse cascade it predicts
\begin{equation}
S_{3}(r)={3 \over 2} \varepsilon_{\alpha} r \;\;\;\;  \mbox{for} \;\;\;\;
r \gg \ell_f
\label{eq:4}
\end{equation}
while for the direct cascade
\begin{equation}
S_{3}(r)={1 \over 8} \eta_{\nu} r^3 \;\;\;\;  \mbox{for} \;\;\;\;
r \ll \ell_f
\label{eq:5}
\end{equation}
The third-order velocity structure function for the simulation $E$ is
shown in Fig.~\ref{fig3}. Both Kolmogorov laws are clearly visible
with the predicted coefficients. We remark that this is the first time that
the two fundamental laws (\ref{eq:4}) and (\ref{eq:5}) are observed
simultaneously.

\begin{figure}[htb!]
\includegraphics[clip=true,keepaspectratio,width=12cm]{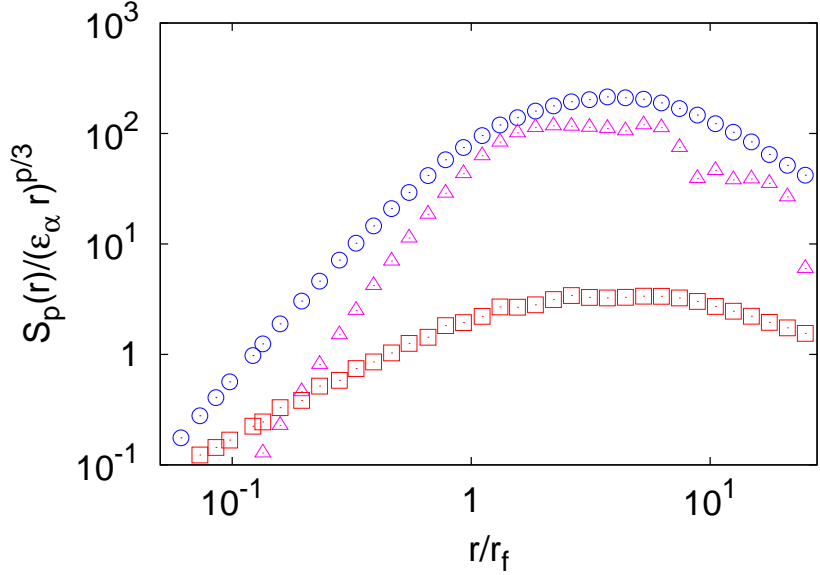}
\caption{(Color online). Longitudinal velocity structure function
$S_{p}(r)$ of order $p=2$ (red squares), $p=4$ (blue circles) and
$p=5$ (pink triangles) from run $E$ and compensated with Kolmogorov-Kraichnan 
prediction $(\varepsilon_{\alpha} r)^{p/3}$.}
\label{fig4}
\end{figure}

In Fig.~\ref{fig4} we plot velocity structure functions of different
orders together with the compensation with Kolmogorov scaling 
$S_{p}(r) \simeq (\varepsilon_{\alpha} r)^{p/3}$. Although the
range of scaling is very small, the presence of a plateau in the 
inverse cascade range of scales confirms that intermittency
corrections are very small or absent in the inverse cascade range 
\cite{bcv_pre00}.

%------------------------------------------------------------------------
%------------------------------------------------------------------------

%For what concerns higher order statistic of the direct cascade 
%one has to consider vorticity increments
%because velocity SF are trivially dominated by the IR contribution.
Velocity structure functions are trivially dominated by the IR contribution 
in the direct cascade range. 
Therefore to investigate higher order statistics of the direct cascade one 
has consider either increments of velocity derivatives (e.g. vorticity 
increments) or velocity second-differences (the latter having the advantage 
of being Reynolds-number-independent in the limit of zero viscosity).  
A Kraichnan energy spectrum $k^{-3}$ would correspond dimensionally
to flat vorticity structure functions, and indeed zero scaling exponents 
for $p>3$ \cite{eyink_physd96} or logarithmic structure functions
\cite{fl_pre94} are predicted in the limit of vanishing viscosity.
Power-law intermittency corrections in the direct cascade
of Navier-Stokes equations are excluded by theory (while logarithmic
corrections are in principle possible), but it is known 
that the presence of a linear friction terms in (\ref{eq:1}) can both 
steepen the spectrum and generate intermittency 
\cite{noag_prl00,bernard_epl00,bcmv_pre02}. 

\begin{figure}[htb!]
\includegraphics[clip=true,keepaspectratio,width=12cm]{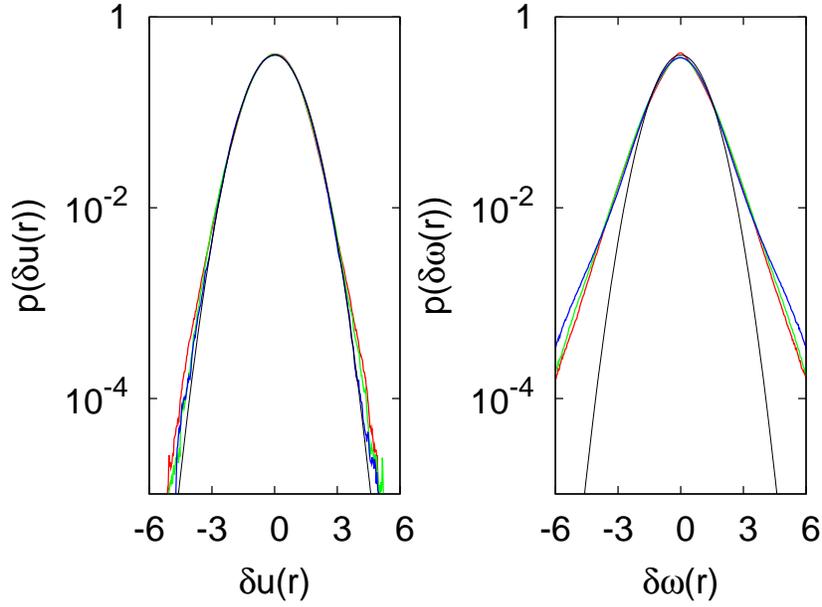}
\caption{(Color online). Probability density functions (pdf) of velocity 
longitudinal increments $\delta u_{\parallel}(r)$ at scales
$r=2.5 \ell_f$ (red outer line) $r=5.0 \ell_f$ (green middle line) and 
$r=10.0 \ell_f$ (blue inner line)
rescaled with rms values (left panel).
Pdf of vorticity increments $\delta \omega(r)$ at scales
$r=0.2 \ell_f$ (red inner line) $r=0.4 \ell_f$ (green middle line) and 
$r=0.8 \ell_f$ (blue outer line)
rescaled with rms values (right panel).
Black curves are standard Gaussian. Data from run $E$.}
\label{fig5}
\end{figure}

We have seen that in our simulations, even at highest
resolution and without friction,
we observe a correction to the spectral exponent, and therefore 
we cannot expect to observe theoretically predicted vorticity 
structure functions. Moreover, because of the limited resolution
of the direct cascade, no clear scaling in vorticity structure 
functions is observed. Nonetheless we can address the issue of 
intermittency by looking at the probability density functions of
fluctuations of vorticity at different scales within the inertial range. 
The result, for run $E$ is shown in Fig.~\ref{fig5} for both
velocity and vorticity increments. 
For what concerns $\delta u_{\parallel}(r)$ we observe self-similar 
pdf in the inertial range of scales, in agreement with the normal
scaling of Fig.~\ref{fig4}. The shape of pdf is very close to Gaussian
with a flatness which is around $3.3$. On the contrary, vorticity increments 
$\delta \omega(r)$ are definitely far from Gaussian with tails which
are found to be very close to exponentials. Nonetheless, the shape of the pdf
does not change substantially in the range of scale of the direct
cascade, an indication of small intermittency also in this case.

%------------------------------------------------------------------------
%------------------------------------------------------------------------

Velocity increments pdf in Fig.~\ref{fig5} cannot be exactly Gaussian as the
energy flux, proportional to $S_3(r)$, requires a positive skewness. 
Energy and enstrophy fluxes are defined in physical space in terms of
filtered fields, as described in \cite{ceewx_prl03,ceerwx_prl06}. 
We introduce a large scale vorticity field 
$\omega_r \equiv G_r \star \omega$ and a large scale velocity
field ${\bf u}_r \equiv G_r \star {\bf u}$ obtained from convolution
with a Gaussian filter $G_r(\bf x)$. From those fields, energy and
enstrophy fluxes $\Pi_r^{(E,Z)}({\bf x},t)$, representing the local transfer
of energy/enstrophy from scales larger than $r$ to scales smaller
to $r$, are defined as
%\begin{equation}
%\Pi_r^{(E)}({\bf x},t)\equiv -(\tau_{\alpha \beta})_r \nabla_r (v_{\beta})_r
%\label{eq:6}
%\end{equation}
%\begin{equation}
%\Pi_r^{(Z)}({\bf x},t)\equiv -(\sigma_{\alpha})_r \nabla_r \omega_r
%\label{eq:7}
%\end{equation}
\begin{equation}
\Pi_r^{(E)}({\bf x},t)\equiv -(\tau_{\alpha \beta})_r \nabla_{\alpha} (v_{\beta})_r
\label{eq:6}
\end{equation}
\begin{equation}
\Pi_r^{(Z)}({\bf x},t)\equiv -(\sigma_{\alpha})_r \nabla_{\alpha} \omega_r
\label{eq:7}
\end{equation}
where 
$(\tau_{\alpha \beta})_r=(v_{\alpha} v_{\beta})_r-(v_{\alpha})_r (v_{\beta})_r$
and 
$(\sigma_{\alpha})_r=(v_{\alpha} \omega)_r-(v_{\alpha})_r \omega_r$.

%------------------------------------------------------------------------
\begin{figure}[htb!]
\includegraphics[clip=true,keepaspectratio,width=12cm]{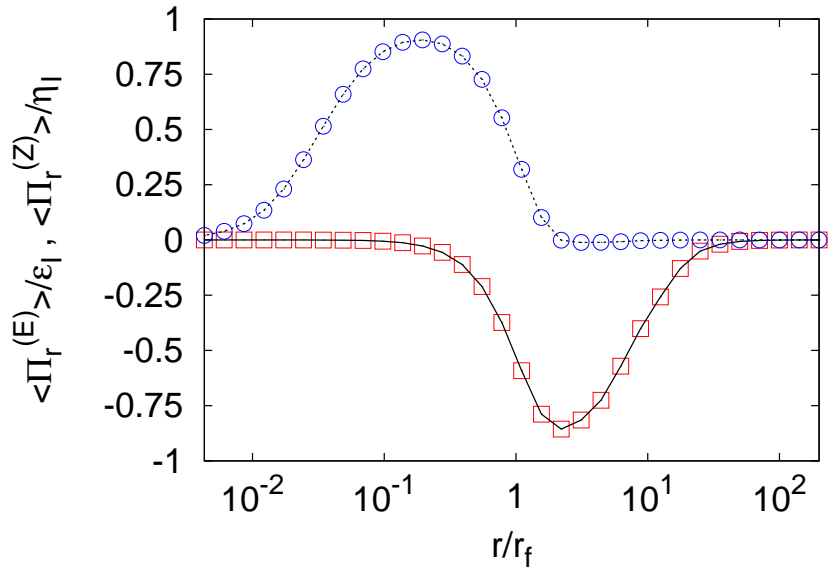}
\caption{(Color online).
Average energy (lower line, red squares) and enstrophy (upper line,
blue circles) fluxes in physical space for run $E$ normalized
with energy/enstrophy inputs.}
\label{fig6}
\end{figure}
%------------------------------------------------------------------------

Figure~\ref{fig6} shows the physical fluxes averaged over space at 
final time of simulation $E$. The two range of scales for the 
energy and enstrophy cascades are evident for $r/\ell_f>1$ and $r/\ell_f<1$
respectively. The finite mean values of fluxes are the results of strong
cancellation: the ratio between the (absolute) mean value and the 
standard deviation at the scales $r_1 \simeq 2.2 \ell_f$ and 
$r_2 \simeq 0.19 \ell_f$ corresponding to the
peaks of the two fluxes are 
$-0.19$ and $0.16$ for energy and enstrophy respectively.
The correlation among the two fluxes is small: the correlation
coefficient between $\Pi_{r_1}^{(E)}$ and $\Pi_{r_2}^{(Z)}$
is only $C(r_1,r_2) \simeq -0.17$ confirming the picture of 
independence of the fluxes in physical spaces already observed at
lower resolution \cite{boffetta_jfm07}.

In conclusion, on the basis of very high resolution numerical
simulations, we obtain strong evidence that the double cascade
theory developed by Kraichnan
more than 40 years ago is substantially correct. This result
required massive resolution as two inertial ranges have
to be resolved simultaneously. It is worth remarking that,
despite some effort \cite{R98,BK05}, the clear observation of the 
two cascade is still lacking in experiments. We hope that
our results will stimulate further experimental investigations
of the double cascade scenario.

%%%%%%%%%%%%%%%%%%%%%%%%%%%%%%%%%%%%%%%%%%%%%%%%%%%%%%%%%%%%%%%%%%%%%%%%%
Numerical simulations has been performed within the 
DEISA Extreme Computing Initiative program ``Turbo2D''.
%%%%%%%%%%%%%%%%%%%%%%%%%%%%%%%%%%%%%%%%%%%%%%%%%%%%%%%%%%%%%%%%%%%%%%%%%
\bibliography{biblio}{}

\end{document}